\documentclass{iau}
\usepackage{graphicx}

\def\be{\begin{equation}}
\def\ee{\end{equation}}
\def\ba{\begin{eqnarray}}
\def\ea{\end{eqnarray}}
\def\go{\mathrel{\raise.3ex\hbox{$>$}\mkern-14mu
             \lower0.6ex\hbox{$\sim$}}}
\def\lo{\mathrel{\raise.3ex\hbox{$<$}\mkern-14mu
             \lower0.6ex\hbox{$\sim$}}}

\def\tomega{\tilde\omega}

\title[High-Frequency QPOs] 
{High-Frequency QPOs and Overstable Oscillations of Black-Hole Accretion Disks}

\author[D. Lai et al.] 
{Dong Lai$^1$, Wen Fu$^{1,2}$, David Tsang$^{1,3,4}$, Jiri Horak$^5$ \and Cong Yu$^{1,6}$}

\affiliation{$^1$ Department of Astronomy, Cornell University, Ithaca, NY 14853, USA
\\email: {\tt dong@astro.cornell.edu} \\
$^2$Department of Physics \& Astronomy, Rice University, Houston, TX, USA \\
$^3$Theoretical Astrophysics, Caltech, Pasadena, CA, USA \\
$^4$Department of Physics \& Astronomy, McGill University, Montreal, Canada\\
$^5$Astronomical Institute of the Academy of Sciences, Prague, CZ \\
$^6$Yunnan Astronomical Observatory, Chinese Academy of Sciences, PRC}

\pubyear{2012}
\volume{290}  
\jname{Feeding compact objects: Accretion on all scales}
\editors{C.M. Zhang, T. Belloni, M. M\'endez \& S.N. Zhang, eds.}

\begin{document}

\maketitle

\begin{abstract}
The physical origin of high-frequency QPOs (HFQPOs) in black-hole
X-ray binaries remains an enigma despite many years of detailed
observational studies. Although there exists a number of models for
HFQPOs, many of these are simply ``notions'' or ``concepts'' without
actual calculation derived from fluid or disk physics. Future progress
requires a combination of numerical simulations and semi-analytic
studies to extract physical insights.  We review recent works on
global oscillation modes in black-hole accretion disks,
and explain how, with the help of general relativistic effects, the
energy stored in the disk differential rotation can be pumped into
global spiral density modes in the disk, making these modes grow to
large amplitudes under certain conditions (``corotational
instability'').  These modes are robust in the presence of disk
magnetic fields and turbulence. The computed oscillation mode
frequencies
are largely consistent with the observed values for HFQPOs in BH X-ray
binaries. The approximate 2:3 frequency ratio is also expected from
this model. The connection of HFQPOs with other disk properties (such
as production of episodic jets) is also discussed.
\end{abstract}

\firstsection 

\section{Introduction}

High-Frequency Quasi-Periodic Oscillations (HFQPOs) in the X-ray
fluxes have been observed from a number of black-hole (BH) X-ray
binaries since the late 1990s. Their frequencies (40-450~Hz) are
comparable to the orbital frequencies at the Innermost Stable Circular
Orbit (ISCO) of BHs with mass $M\sim 10 M_{\odot}$, and do not vary
significantly in response to sizeable (factors of 3-5) changes in the
luminosity.  Compared to the low-frequency QPOs, the HFQPOs have low
amplitudes ($0.5-2\%$ rms at 2-60~keV) and low coherence ($Q\sim
2-10$), and are only observed in the ``intermediate state'' (or
``steep power law state'') of the X-ray binaries (for reviews, see
Remillard \& McClintock 2006; Belloni et al.~2012).
In addition, several systems show pairs of QPOs with frequency ratios
close to $2:3$.  It is generally recognized that HFQPOs may provide
important information about the BH (mass and spin) and about the
dynamics of inner-most accretion flows.

{\bf Current status of QPO Theory/Models:}
The physical origin of HFQPOs is currently unclear. 
Below we comment on some of the ideas/models that have been suggested.

(i) In the spot models (Stella et al.~1999),
the HFQPOs arise from the Doppler modulation of the radiation from
spots orbiting in the inner part of the BH accretion disk. It is
not clear how the position of the spot (a free parameter in these
models) is determined and how the spot can survive the differential
rotation of the disk. 

(ii) Another class of models identify HFQPOs with
various oscillation modes of a finite, pressure-supported accretion
torus (Rezzolla et al.~2003; Blaes et al.~2006).  It is not
clear that the accretion flow can be approximated by a torus and how
the position and size of the torus are determined.  

(iii) The harmonic relation between the observed frequencies led
Abramowicz \& Kluzniak (2001) to suggest that the HFQPOs are a result
of a nonlinear resonance.  However, so far no fluid dynamical model
producing these resonances has been developed.
Detailed calculations of the resonant
coupling between the epicyclic modes in slender tori indicates that
such resonance is very weak (Hor\'ak 2008).

(iv) A large class of theoretical models is based on the relativistic
diskoseismology (Kato 2001).
The $g$-modes (also called inertial-gravity modes) have
attracted most attention because their existence does not require a
reflective inner disk boundary.  These oscillations have at least one
node in the vertical direction and their restoring force results from
rotation and gravity. Unfortunately, the non-axisymmetric $g$-modes
are either damped due to corotation resonance (Kato 2003; Li et al.~2003)
or have frequencies too high compared to HFQPOs.
The axisymmetric $g$-mode trapped around the maximum of the radial
epicyclic frequency may account for some single HFQPO frequencies (see
Wagoner 2012), and several studies suggested that they may be
resonantly excited in the warped or eccentric disks (Kato 2008;
Ferreira \& Ogilvie 2008).
However, the self-trapping property of $g$-modes can be easily
destroyed by a weak (sub-thermal) disk magnetic field (Fu \& Lai 2009)
and turbulence (Arras et al.~2006; Reynolds \& Miller 2009).

Related to (iv) above, perhaps more promising is the disk $p$-modes.
In the following, we review recent works on disk $p$-modes
from the Cornell group and other related efforts
(see Lai \& Tsang 2009, Tsang \& Lai 2008,2009a,b,c; Fu \& Lai 2009,2011a,b,2012a,b;
Horak \& Lai 2012).

\section{Disk P-modes: Corotational Instability}

\begin{figure}[b]
 \vspace*{-0.8 cm}
\begin{center}
\includegraphics[width=5.4in]{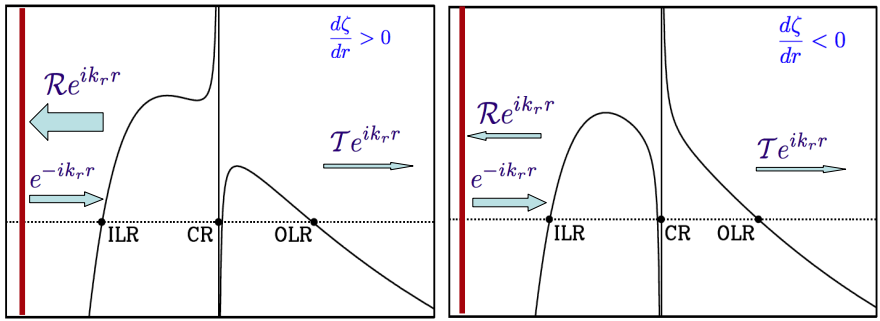}
\vspace*{-5.1 cm}
\caption{\footnotesize
Super-reflection and corotational instability of nonaxisymmetric waves
in rotating disks. The solid curves depict the effective potential for
wave propagation: Density waves can propagate either inside the 
ILR
or outside the OLR.
Between the ILR and OLR lies the corotation resonance (CR) and
the associated Rossby zone (where the effective potential is
negative).  An incident wave $e^{-ik_r r}$ propagating from small
radii towards the corotation barrier will be partially reflected
(${\cal R}e^{ik_r r}$) and partially transmitted (${\cal T}e^{ik_rr}$).
Since the waves inside CR ($r<r_{\rm CR}$) carry negative energies,
while those outside CR carry positive energies, we have
$(-1)=(-1)|{\cal R}|^2+|{\cal T}|^2+{\cal D}_c$ from energy conservation,
or $|{\cal R}|^2=1+|{\cal T}|^2+{\cal D}_c$, where ${\cal D}_c$ is the wave
energy absorption at CR.  When the background disk has $d\zeta/dr>0$
at CR [where $\zeta=\kappa^2/(2\Omega\Sigma)$ is the vortensity], the
Rossby zone lies just outside CR, thus ${\cal D}_c>0$ and $|{\cal
  R}|^2\simeq 1+{\cal D}_c>1$, we have super-reflection (the left
panel). If the disk has a reflective inner boundary (represented by
the red vertical line), then the stationary waves (modes) trapped
between $r_{\rm in}$ and ILR will be overstable (with growing
amplitude). When the background disk has $d\zeta/dr<0$, there is no
super-reflection ($|{\cal R}|<1$) and the modes are damped by the
corotational wave absorption (the right panel).}
   \label{fig1}
\end{center}
\end{figure}

Disk $p$-modes (also called inertial acoustic modes) represent nearly
horizontal oscillations with almost no vertical structure.
They are trapped between the inner boundary of the disk and 
the inner Lindblad resonance (ILR; see below)
Because of their simple 2D structure,
the basic wave properties of the p-modes (e.g., propagation diagram)
are not strongly affected by disk magnetic fields (Fu \& Lai 2009) and
are likely robust in the presence of disk turbulence.
Lai \& Tsang (2009) (see also Tsang \& Lai 2008,2009c) showed that the
non-axisymmetric $p$-modes can naturally grow due to the
\textit{corotational instability}. This can be understood as follows.

Consider a 2D fluid disks without self-gravity
and magnetic field. For the disturbances of the form $e^{im\phi-i\omega t}$,
where $m>0$ and $\omega$ is the wave (angular) frequency,
the WKB dispersion relation for density waves reads
\be
\tomega^2=\kappa^2+k_r^2c_s^2,\qquad\qquad ({\rm where}~~\tomega\equiv\omega-
m\Omega)
\label{eq:disp}
\ee
where $\Omega(r)$ is the disk rotation frequency,
$\kappa(r)$ is the radial epicyclic frequency, $k_r$ is the radial
wavenumber, and $c_s$ is the sound speed. Thus waves can propagate
either inside the Inner Lindblad resonance (ILR, defined by
$\tomega=-\kappa$) or outside the Outer Lindblad resonance (OLR,
$\tomega=\kappa$). The region between ILR and OLR is evanescent
(``corotation barrier'') except for a narrow ``Rossby zone'' near the
corotation resonance (CR, where $\tomega=0$).
The significance of CR is that waves can
be absorbed by the background flow since the wave patten speed
$\omega/m$ matches the flow rotation rate $\Omega$ at CR. Depending on the
sign of $d\zeta/dr$, where $\zeta\equiv\kappa^2/(2\Omega\Sigma)$ is
the \underline{\it vortensity} of the background flow ($\Sigma$ is the surface 
density of the disk), the Rossby zone can lie just inside or outside CR, and the
corotational wave absorption can be either positive or negative (see
Fig.~1).

This corotational instability of non-axisymmetric $p$-modes arises
because of two effects: (1) Since the waves inside the ILR carry
negative energies while those outside the OLR
carry positive energies, the
leakage of the p-waves through the corotation barrier (between ILR and
OLR) leads to mode growth. (2) More importantly, when the vortensity
of the disk flow, $\zeta=\kappa^2/(2\Omega\Sigma)$ 
[this applies to barotropic fluid, for which the pressure is a unique
  function of the density; see Tsang \& Lai (2009c) for the
  non-barotropic case], has a positive slope at the corotation radius
wave absorption at the corotation resonance leads to amplification of
the trapped p-mode.

{\bf General Relativity} plays an important role in the corotational
instability of p-modes. In Newtonian theory, $\kappa=\Omega$, so
$d\zeta/dr<0$ when $\Sigma$ does not vary strongly and hence the mode cannot
grow.  In GR, $\kappa$ reaches a maximum at a finite radius before
dropping to zero at $r_{\rm ISCO}$. Thus, if the mode frequency is such that
$d\zeta/dr>0$ at corotation, then the mode can grow.

\section{Overstable P-modes and HFQPOs}

Our linear calculations and nonlinear simulations (based on
pseudo-Newtonian potential) have demonstrated
the corotational instability of disk p-modes (Lai \& Tsang 2009;
Tsang \& Lai 2009c; Fu \& Lai 2011a,2012a). In particular, the
low-order p-modes (trapped between the inner disk edge and the ILR)
have the following properties:

(i) The mode frequency is $\omega\simeq \beta m\Omega_{\rm in}$,
where $\beta=0.5$-$0.75$ (depending on the disk models and the inner 
boundary condition. The mode frequency varies (by about $10\%$) as
$\dot M$ changes by a factor of 3. The frequency is consistent with the
estimated BH mass and spin of X-ray binaries.
The frequency ratio is approximately $1:2:3,\cdots$, but not exactly. 

(ii) The mode growth rate (due to corotation resonance) is much smaller
for $m=1$ than for $m=2,3,4,\cdots$. Since high-$m$ modes are hard to observe,
it is reasonable to expect that the most visible modes are $m=2,3$.

(iii) The nonlinear mode frequency is close to the linear mode frequency
(Fu \& Lai 2012b).

\section{Issues, Complications and Prospects}

All the above make p-modes a promising candidate for explaining HFQPOs
in X-ray binaries.  However, there are a number of issues and complications
that must be considered:

(i) {\bf Mode Damping due to Radial Infall:} For standard thin disks,
the radial infall velocity increases rapidly near $r_{\rm ISCO}$ --
this can be described in the slim disk model. Such infall tends to
damp the p-modes (but not completely, because of the sharp density
gradient around $r_{\rm ISCO}$; see Lai \& Tsang 2009).  So we have a
competition between mode growth (due to corotation resonance) and
damping (due to infall). The net result is that for standard thin
disks, p-modes are likely damped. This is consistent with the
observations that no HFQPOS are detected in the thermal (low-soft)
state of BH X-ray binaries.

(ii) {\bf Effects of magnetic fields}: However, real disks
(particularly those in the ``intermediate state'' when HFQPOs are
observed) are more complicated, and have magnetic fields. As noted
above, the p-mode frequencies are only slightly affected by the
magnetic field, provided that $r_{\rm in}$ is close to $r_{\rm ISCO}$.
However, the mode growth rates can be significant modified by the
$B$-field. There are three effects: (1) Toroidal fields inside the
disk tend to surpress the corotational instability (Fu \& Lai
2011a,b).  (2) Magnetic fields can accumulate inside $r_{\rm ISCO}$,
forming a magnetosphere around the BH. Thus, the inner disk edge may
be more reflective than the standard thin disk (Tsang \& Lai 2009b; Fu
\& Lai 2012a). This can enhance the net growth rate of p-modes.  (3)
Large-scale poloidal magnetic fields threading the disk can provide a
coupling between the disk and the disk magnetosphere (Yu \& Lai 2013,
in prep; see Yu \& Lai 2012 for a related problem).  This tends to
enhance the corotational instability and increase the growth rate of
p-modes (see also Tagger \& Varniere 2006).

The last two points may be of particular importance. Large-scale
magnetic fields provide a natural means to launch jets/outflows from
accretion disks.  Magnetic fields threading the BH can also lead to
jets from the rotating BHs (McKinney et al.~2012). In BH x-ray
binaries, episodic (ballistic) jets are observed in the intermediate
state, and this is the same state during which HFQPOs are detected.
In our model, the disk and corona (coupled by a large-scale poloidal 
$B$-field) oscillate together, with
the ``clock'' mainly set by the disk dynamics. The corona is
important, however, since it can lead to the variation of hard x-ray
photon flux as observed in HFQPOs.

The current understanding of accretion flows in the intermediate state
is rather poor (Done et al.~2007). It is even not clear that 
$r_{\rm in}$ extends to $r_{\rm ISCO}$ in this state. 
We suggest that in the intermediate
state, large-scale $B$-fields are temporarily/episodically created
(e.g., magnetic fields buoyantly rise from the collapsed ADAF state
and followed by reconnection).  This will then allow the p-modes to
grow and manifest as HFQPOs; at the same time, it will lead to the
production of episodic jets.

(iii) {\bf Full GR Calculations}. Horak \& Lai (2012) presented full
GR calculations of overstable p-modes of BH accretion disks. In
particular, they derived the GR version of the corotational
instability criterion and generalized the Newtonian disk
vortensity. The results are qualitatively in agreement with the
pseudo-Newtonian results discussed above, with one important
difference: For Kerr BHs with the spin parameter close to unity, the
mode frequency is close to $m\Omega_{\rm in}$ (i.e., $\beta\simeq
1$). Assuming $r_{\rm in}=r_{\rm ISCO}$, this would lead to a 
tension between the simplest p-mode theory and the observational
constraints on the BH mass and spin.  The discrepency becomes most
severe for GRS 1915+105 ($M=14\pm 4.4M_\odot$), which may have two
pairs of HFQPOs (41 and 67~Hz, 113 and 168~Hz), and whose spin
parameter has been constrained to be $a>0.975$ using the continuum
fitting method (see Narayan \& McClintock 2012).
Thus, it appears that for the simplest disk models,
non-axisymmetric p-modes have frequencies that are too high compared
to the observation of HFQPOs in BH X-ray binaries (assuming the BH
parameters obtained using the continum fitting method is reliable), at
least for some systems. Several effects may decrease the theoretical
p-mode frequencies. For example, a higher disk sound speed leads to
lower mode frequencies, and a steeper surface density profile (larger
$p$ in $\Sigma\propto r^{-p}$) reduces $\omega_{\rm min}$, the minimum
frequency for the corotational wave absorption to amplify the mode.
Magnetic fields may also play an important role [see (ii) above]. In
particular, our calculations so far assume that the inner disk radius
coincides with the ISCO. This may not be the case during the
``intermediate state'' (Done et al.~2007; Oda et al.~2010).  As noted
before, a major uncertainty in calculating the disk p-modes is the
inner disk boundary condition. When magnetic fields advect inwards in
the accretion disk and accumulate around the BH, (Bisnovatyi-Kogan \&
Ruzmaikin 1974; Igumenshchev et al.~2003; Rothstein \& Lovelace 2008;
Guilet \& Ogilvie 2012), the inner disk radius $r_{\rm in}$ may be larger than 
$r_{\rm ISCO}$. This reduces $\Omega_{\rm in}$ relative to $\Omega_{\rm ISCO}$, 
leading to lower p-mode frequencies.

(iv) {\bf Numerical Simulations.}
Wen \& Lai (2012b) carried out 2D inviscid hydrodynamic simulations of
overstable p-modes in BH accretion discs assuming that the 
disk inner boundary is sufficiently reflective. 
The simulations confirmed the linear growth of p-modes and showed that the
mode growth saturates when the radial velocity perturbation becomes
comparable to the disc sound speed. During the saturation stage, the
primary disc oscillation frequency differs only slightly (by less than
a few percent) from the linear mode frequency. Sharp features in the
fluid velocity profiles at this stage suggest that the saturation results
from nonlinear wave steepening and mode-mode interactions.

Such 2D simulations, while useful, do not capture various complexities
(e.g., magnetic field, turbulence, radiation) associated with real BH
accretion discs. Numerical MHD simulations (including GRMHD) are
playing an increasingly important role in unraveling the nature of BH
accretion flows. A number of ``numerical QPOs'' have been reported in
the literature. For example, Henisey et al.~(2009,2012) found evidence
of excitation of wave modes in simulations of tilted BH accretion
disks.  Hydrodynamic simulations using $\alpha$-viscosity (Chan 2009;
O'Neill et al.~2009) showed wave generation in the inner disk region
by viscous instability (Kato 2001). The MHD simulations by O'Neill et
al.~(2011) revealed possible LFQPOs due to disk dynamo cycles.
Dolence et al.~(2012) reported transient QPOs in the numerical models
of radiatively inefficent flows for Sgr A$^\star$. McKinney et
al.~(2012) found QPO signatures associated with the interface between
the disc inflow and the bulging jet magnetosphere (see Fu \& Lai
2012a). It is not clear if these QPOs are directly comparable with the
observations of BH X-ray binaries. Obviously, much work remains to be
done to understand the physics of HFQPOs. A combination of full
numerical simuations and semi-analytic works (together with
``controlled'' numerical experiments), as well as future high signal-to-noise
observations of HFQPOs (e.g., with LOFT; Feroci et al.~2012), are 
needed to make progress in this field.

{\bf Acknowledgements}:
This work has been supported in part by NSF grants AST-1008245 and
AST-1211061, and NASA grants NNX12AF85G and NNX10AP19G.


\begin{thebibliography}

\bibitem[\protect\citeauthoryear{AL}{1976a}]{AL76a}
Abramowicz M. A., Kluzniak, W. 2001, A\&A, 374, L19

\bibitem[\protect\citeauthoryear{AL}{1976a}]{AL76a}
Arras P., Blaes O.M., Turner N. J., 2006, ApJ, 645, L65

\bibitem[\protect\citeauthoryear{AL}{1980}]{AL80}
Belloni, T.M., Sanna, A., \& Mendez, M. 2012, MNRAS, in press (axrXiv:1207.2311)

\bibitem[\protect\citeauthoryear{BR}{1974}]{BR74}
Bisnovatyi-Kogan G. S., Ruzmaikin A. A., 1974, Ap\&SS, 28, 45


\bibitem[\protect\citeauthoryear{Chan}{2009}]{Chan09}
Chan C.-K., 2009, ApJ, 704, 68

\bibitem[\protect\citeauthoryear{DGK}{2007}]{DGK12}
Dolence, J.C., Gammie, C.F., Shiokawa, H., Noble, S.C. 2012, ApJ, 746, L10

\bibitem[\protect\citeauthoryear{Chandrasekhar}{1961}]{C61}
Done C., Gierlinski M., Kubota A., 2007, Astron. Astrophys. Rev., 15, 1

\bibitem[\protect\citeauthoryear{FuLai}{2009}]{FuLai09}
Ferreira B. T., Ogilvie G. I., 2008, MNRAS, 386, 2297

\bibitem[\protect\citeauthoryear{FuLai}{2009}]{FuLai09}
Feroci, M., et al. 2012, arXiv:1209.1497

\bibitem[\protect\citeauthoryear{FuLai}{2009}]{FuLai09}
Fu W., Lai D., 2009, ApJ, 690, 1386

\bibitem[\protect\citeauthoryear{FuLai}{2011}]{FuLai2011a}
Fu W., Lai D., 2011a, MNRAS, 410, 399

\bibitem[\protect\citeauthoryear{FuLai}{2011}]{FuLai2011a}
Fu W., Lai D., 2011b, MNRAS, 410, 1617

\bibitem[\protect\citeauthoryear{FuLai}{2012}]{FuLai2012}
Fu W., Lai D., 2012a, MNRAS, 423, 831

\bibitem[\protect\citeauthoryear{FuLai}{2012}]{FuLai2012}
Fu W., Lai D., 2012b, MNRAS, submitted (arXiv:1212.2215)



\bibitem[\protect\citeauthoryear{FuLai}{2012}]{FuLai2012}
Guilet, J., Ogilvie, G.I. 2012, MNRAS, in press (arXiv:1212.0855)

\bibitem[\protect\citeauthoryear{HBFF}{2009}]{HBFF09}
Henisey K. B., Blaes O. M., Fragile P. C., Ferreira B. T., 2009, ApJ, 706, 705

\bibitem[\protect\citeauthoryear{HBFF}{2012}]{HBFF12}
Henisey K. B., Blaes O. M., Fragile P. C., 2012, ApJ, 761, 18

\bibitem[\protect\citeauthoryear{INA}{2003}]{INA03}
Hor\'ak, J. 2008, A\&A, 486, 1

\bibitem[\protect\citeauthoryear{INA}{2003}]{INA03}
Hor\'ak, J., \& Lai, D. 2012, MNRAS, submitted

\bibitem[\protect\citeauthoryear{INA}{2003}]{INA03}
Igumenshchev I. V., Narayan R., Abramowicz M. A., 2003, ApJ, 592, 1042

\bibitem[\protect\citeauthoryear{Kato}{2001}]{Kato01}
Kato S., 2001, PASJ, 53, 1

\bibitem[\protect\citeauthoryear{Kato}{2003}]{Kato03a}
Kato S., 2003, PASJ, 55, 257

\bibitem[\protect\citeauthoryear{Kato}{2008}]{Kato08}
Kato S., 2008, PASJ, 60, 111



\bibitem[\protect\citeauthoryear{LT09}{2009}]{LT09}
Lai D., Tsang D., 2009, MNRAS, 393, 979

\bibitem[\protect\citeauthoryear{LN04}{2004}]{LN04}
Li L., Goodman J., Narayan R., 2003, ApJ, 593, 980


\bibitem[\protect\citeauthoryear{LS}{1995}]{LS95}
McKinney, J.C., Tchekovskoy, A., Blandford, R.D. 2012, MNRAS, 423, 3083

\bibitem[\protect\citeauthoryear{LS}{1995}]{LS95}
Meheut, H., Yu, C., Lai, D. 2012, MNRAS, 422, 2399


\bibitem[\protect\citeauthoryear{LS}{1995}]{LS95}
Narayan, R., Goldreich, P., Goodman, J. 1987, MNRAS, 228, 1

\bibitem[\protect\citeauthoryear{LS}{1995}]{LS95}
Narayan, R., McClintock, J. 2012, MNRAS, 419, L69

\bibitem[\protect\citeauthoryear{Nowak}{1991}]{NW91}
Oda H., et al., 2010, ApJ, 712, 639


\bibitem[\protect\citeauthoryear{ORM}{2009}]{ORM09}
O'Neill S. M., Reynolds C. S., Miller C. M., 2009, ApJ, 693, 1100

\bibitem[\protect\citeauthoryear{ORMS}{2011}]{ORMS11}
O'Neill S. M., Reynolds C. S., Miller C. M., Sorathia K., 2011, ApJ, 736, 107


\bibitem[\protect\citeauthoryear{RM}{2006}]{RM06}
Remillard R. A., McClintock J. E., 2006, ARA\&A, 44, 49

\bibitem[\protect\citeauthoryear{RM}{2009}]{RM09}
Reynolds C. S., Miller M. C., 2009, ApJ, 692, 869


\bibitem[\protect\citeauthoryear{RL}{2008}]{RL08}
Rothstein D. M., Lovelace R. V. E., 2008, ApJ, 677, 1221

\bibitem[\protect\citeauthoryear{SSL}{1995}]{SSL95}
Stella L., Vietri M., Morsink S. M. 1999, ApJ, 524, L63



\bibitem[\protect\citeauthoryear{Tsang}{2008}]{TL08}
Tagger M., Varniere P. 2006, ApJ, 652, 1457

\bibitem[\protect\citeauthoryear{Tsang}{2008}]{TL08}
Tsang D., Lai D., 2008, MNRAS, 387, 446

\bibitem[\protect\citeauthoryear{Tsang}{2009}]{TL09}
Tsang D., Lai D., 2009a, MNRAS, 393, 992

\bibitem[\protect\citeauthoryear{Tsang}{2009}]{TL09}
Tsang D., Lai D., 2009b, MNRAS, 396, 589

\bibitem[\protect\citeauthoryear{Tsang}{2009}]{TL09}
Tsang D., Lai D., 2009c, MNRAS, 400, 470




\bibitem[\protect\citeauthoryear{Wagoner}{1999}]{Wagoner99}
Wagoner, R.V. 2012, ApJ, 752, L18

\bibitem[\protect\citeauthoryear{yu}{1999}]{yu12}
Yu, C., Lai, D. 2012, MNRAS, in press (arXiv:1212.1219)


\end{thebibliography}
\end{document}